\documentclass[pra,twocolumn,superscriptaddress,nofootinbib,showpacs,a4paper]{revtex4}
\usepackage{amsmath}
\usepackage{amsfonts}
\usepackage{graphicx}

\begin{document}
\title{Optimal measurements for relative quantum information}
\author{Stephen D. Bartlett}
\email{bartlett@physics.uq.edu.au}
\affiliation{School of Physical Sciences, The University of Queensland,
  Queensland 4072, Australia}
\author{Terry Rudolph}
\email{t.rudolph@imperial.ac.uk}
\affiliation{Optics Section, Blackett Laboratory, Imperial College
  London, London SW7 2BZ, United Kingdom}
\author{Robert W. Spekkens}
\email{rspekkens@perimeterinstitute.ca}
\affiliation{Perimeter Institute for Theoretical Physics, Waterloo,
  Ontario N2J 2W9, Canada}
\date{6 July 2004}

\begin{abstract}
  We provide optimal measurement schemes for estimating relative
  parameters of the quantum state of a pair of spin systems.  We prove
  that the optimal measurements are joint measurements on the pair of
  systems, meaning that they cannot be achieved by local operations
  and classical communication.  We also demonstrate that in the limit
  where one of the spins becomes macroscopic, our results reproduce
  those that are obtained by treating that spin as a classical
  reference direction.
\end{abstract}
\maketitle

\section{Introduction}

Whenever a system can be decomposed into parts, a distinction can be
made between collective and relative degrees of freedom.  Collective
degrees of freedom describe the system's relation to something
external to it, while the relative ones describe the relations between
its parts.  Encoding information into collective degrees of freedom is
problematic in situations where the parts of the system are subject to
an environmental interaction that does not distinguish them
(collective decoherence), or if the external reference frame (RF) with
respect to which they were prepared is unknown, or if a superselection
rule applies to the total system~\cite{Ver03,BW03,KMP04}.  In contrast,
encoding information preferentially into the relative degrees of
freedom has been shown to offer advantages in these situations, with
applications in quantum computation~\cite{Zan97,Kwi00},
communication~\cite{Bar03,Bou03} and cryptography~\cite{Wal03,Boi03}.

If the relative encoding is not perfect or is itself subject to some
noise, it becomes important to identify measurement schemes for
estimating relative parameters.  Such measurements have been discussed
recently in connection with their ability to induce a relation between
quantum systems that had no relation prior to the measurement, e.g.,
inducing a relative phase between two Fock states~\cite{Jav96,San03}
or a relative position between two momentum eigenstates~\cite{Rau03}.
Also, measurements of relative parameters are critical for achieving
\emph{programmable} quantum measurements~\cite{Dus02,Fiu02}.

Schemes for estimating relative quantum information are also
interesting in their own right.  They include such natural tasks as
estimating the distance between two massive particles, the phase
between two modes of an electromagnetic field, or the angle between a
pair of spins.  In this paper, we focus on this last example: optimal
relative parameter estimation for a rotational degree of freedom given
a pair of spin systems.  Note that our problem is complementary to
that of determining the optimal measurement schemes for estimating
\emph{collective} parameters for a rotational degree of freedom, a
subject of many recent investigations~\cite{Per01,Bag01,Chi04}.

One scheme for estimating such relative parameters is to measure each
system independently with respect to an external RF, e.g., to perform
an optimal estimation of each spin direction and then to calculate the
angle between these estimates.  We prove that any such \emph{local}
scheme, performed using only local operations and classical
communication (LOCC), cannot be optimal; the ability to perform
\textit{joint} measurements is necessary to achieve the
optimum~\cite{Mas95}.  We also prove that the optimal measurement for
estimating a relative angle can be chosen to be
rotationally-invariant, demonstrating that an external RF is not
required.  We investigate the information gain that can be achieved as
different aspects of the estimation task are varied, such as the prior
over the relative angle or the magnitude of the spins.

Previous studies into parameter estimation have not considered the
role of the RF (implicitly presumed to be classical); our development
of relative parameter estimation is appropriate for the case where the
RF is itself quantized.
We investigate quantum-classical correspondence of RFs by considering
the limit in which one of the spins becomes large, and demonstrate
that our optimal relative measurement yields the same information gain
in this limit as does the optimal measurement for estimating a spin's
direction relative to a classical RF.  Interestingly, we also find
that the need for joint measurements disappears in this limit.  These
results contribute to our understanding of how collective degrees of
freedom, which are defined with respect to a classical RF, can be
treated as relative ones between quantized systems.  Such an
understanding is likely to be critical for quantum gravity and
cosmology, wherein all degrees of freedom are expected to be
relative~\cite{Rov98}.

\section{Relative parameter estimation}

Consider states in the joint Hilbert space $\mathbb{H}_{j_1}\otimes
\mathbb{H}_{j_2}$ of a spin-$j_1$ and a spin-$j_2$ system.  This
Hilbert space carries a collective tensor representation $R(\Omega) =
R_{j_1}(\Omega)\otimes R_{j_2}(\Omega)$ of a rotation $\Omega \in$
SU(2) where each system is rotated by the same amount.  We can
parametrise the states in $\mathbb{H}_{j_1}\otimes \mathbb{H}_{j_2}$
by two sets of parameters, $\alpha$ and $\Omega$, such that a state
$\rho_{\alpha,\Omega}$ transforms under a collective rotation as
\begin{equation}
  \label{eq:GlobalParameter}
  R(\Omega') \rho_{\alpha,\Omega}
  R(\Omega')^\dag = \rho_{\alpha,\Omega' \Omega} \, .
\end{equation}
Defining a collective parameter as one whose variation corresponds to
a collective rotation of the state, and a relative parameter as one
that is invariant under such a rotation, we see that $\alpha$ is
relative and $\Omega$ is collective.  

Note that a typical parameter will be neither collective nor relative.
However, in situations where a superselection rule for the group of
collective transformations applies, or when all systems that can serve
as a classical RF for the collective degrees of freedom have been
quantized, one finds that all collective parameters become
operationally meaningless, and \emph{all} observable parameters are
relative.

In the special case of two spins each prepared in an SU(2) coherent
state~\cite{Arr72} (discussed below), there is only a single relative
parameter: the angle between the two spins.  Note that this angle
cannot be perfectly determined by a single measurement (there exist
sets of states with different values of this relative parameter that
are nonorthogonal) and thus we refer to information about this
relative parameter as quantum.

Suppose that Alice prepares a pair of spins in the state
$\rho_{\alpha,\Omega}$ and Bob wishes to acquire information about the
relative parameters $\alpha$ without having any prior knowledge of the
collective parameters $\Omega$. The most general measurement that can be
performed by Bob is a positive operator valued measure
(POVM)~\cite{Per95} represented by a set of operators $\{E_\lambda\}$.
Upon obtaining the outcome $\lambda$, Bob uses Bayes' theorem to update
his knowledge about $\alpha,\Omega$ from his prior distribution
$p(\alpha,\Omega)$, to his posterior distribution: 
\begin{equation}
  \label{eq:Posterior}
  p(\alpha,\Omega
  |\lambda )=\frac{ \mathrm{Tr} (E _{\lambda }\rho _{\alpha,\Omega})
  p(\alpha,\Omega )}{p(\lambda)} \,,
\end{equation}
where 
\begin{equation}
  \label{eq:PriorOfLambda}
  p(\lambda)=\int \mathrm{Tr}(E _{\lambda }\rho _{\alpha,\Omega})
  p(\alpha,\Omega )\, \mathrm{d}\alpha\, \mathrm{d}\Omega \, .
\end{equation}
Assuming that Bob has no prior knowledge of $\Omega$, we may take
$p(\alpha,\Omega)\, \mathrm{d}\alpha \, \mathrm{d}\Omega =
p(\alpha)\,\mathrm{d}\alpha\, \mathrm{d}\Omega$ where $p(\alpha)$ is Bob's
prior probability density over $\alpha$ and $\mathrm{d}\Omega$ is the
SU(2) invariant measure.

Any measure of Bob's information gain about $\alpha$ can depend only
on the prior and the posterior distributions over $\alpha$ for every
$\lambda$.  The latter are obtained by marginalization of the
$p(\alpha,\Omega |\lambda )$, and are given by 
\begin{equation}
  \label{eq:Marginalize}
  p(\alpha |\lambda) =
  \frac{\mathrm{Tr} (E _{\lambda} \rho _{\alpha}) p(\alpha )}{p(\lambda)}\,,
\end{equation}
where 
\begin{equation}
  \rho_{\alpha}=\int
  R(\Omega')\rho_{\alpha,\Omega}R(\Omega')^{\dag} \mathrm{d}\Omega' \,.
\end{equation}
For a given POVM $\{E_\lambda\}$, note that any other POVM related by
a collective rotation (i.e., $E'_{\lambda}=R(\Omega )E_{\lambda }R(\Omega
)^{\dag }$) yields precisely the same posterior distributions over
$\alpha$.  This property also holds true for the POVM with elements
$\bar{E}_\lambda = \int R(\Omega)E_\lambda R(\Omega)^{\dag}
\mathrm{d}\Omega$, which is rotationally-invariant, that is,
\begin{equation}
  R(\Omega) \bar{E} _{\lambda} R(\Omega )^{\dag } =
  \bar{E}_{\lambda }\,, \quad \forall\ \Omega \in \mathrm{SU(2)}\,.
  \label{eq:RotInvariantPOVM}
\end{equation} 
We define POVMs that yield the same posterior distribution over
$\alpha$ to be \emph{informationally equivalent}.  For every POVM,
there exists a rotationally-invariant POVM of the
form~(\ref{eq:RotInvariantPOVM}) that is informationally equivalent,
and thus it is sufficient to consider only rotationally-invariant
POVMs in optimizing Bob's choice of measurement.  These can be
implemented without an external RF for spatial orientation.  Moreover,
they have a very particular form, as we now demonstrate.

The joint Hilbert space for the two spins decomposes into a
multiplicity-free direct sum of irreducible representations (irreps)
of SU(2), i.e., eigenspaces $\mathbb{H}_{J}$ of total angular momentum
$J$.  Using Schur's lemma~\cite{Ste95}, it can be shown that any
positive operator satisfying~(\ref{eq:RotInvariantPOVM}) can be
expressed as a positive-weighted sum of projectors $\Pi _{J}$ onto the
subspaces $\mathbb{H}_{J}$, that is, as $E _{\lambda
}=\sum_{J}s_{\lambda , J}\Pi _{J}$, where $s_{\lambda ,J}\ge 0$.  In
order to ensure that $\sum_{\lambda }E _{\lambda }=\mathbb{I}$, we
require that $\sum_{\lambda}s_{\lambda ,J}=1$, so that $s_{\lambda ,
  J}$ is a probability distribution over $\lambda$.  The $\{E_{\lambda
}\}$ can be obtained by random sampling of the projective measurement
elements $\{\Pi _{J}\}$, and such a sampling cannot increase the
information about the relative parameters (quantified by some concave
function such as the average information gain).  Thus, the most
informative rotationally-invariant POVM is simply the projective
measurement $\{\Pi _{J}\}$.

We have proved the main result of the paper, which can be summarized
as follows: If the prior over collective rotations $\Omega$ is
uniform, then for \emph{any} prior over the relative parameters
$\alpha$, the maximum information gain (by any measure) can be
achieved using the rotationally-invariant projective measurement
$\{\Pi_J\}$.

A useful way to understand this result is to note that our estimation
task is equivalent to one wherein Alice prepares a state
$\rho_{\alpha}$ (rather than $\rho_{\alpha,\Omega}$).  Because the
$\rho_{\alpha}$ are rotationally-invariant, they are also positive
sums of the $\Pi_J$ and thus may be treated as classical probability
distributions over $J$.  The problem reduces to a discrimination among
such distributions, for which Bob can do no better than to measure the
value of $J$.

We now apply this result to several important and illustrative
examples of relative parameter estimation.  We shall quantify the
degree of success in the estimation by the average decrease in Shannon
entropy of the distribution over $\alpha$~\cite{Per95}, which is
equivalent to the average (Kullback-Leibler) relative information
between the posterior and the prior distributions over $\alpha$,
specifically $I_{\rm av}=\sum_{\lambda}p(\lambda)I_{\lambda}$, where
\begin{equation}
  \label{eq:Ilambda}
  I_{\lambda}=\int
  p(\alpha|\lambda)\log_2\bigl[p(\alpha|\lambda)/p(\alpha)\bigr] {\rm
  d}\alpha \,.
\end{equation}
We refer to this quantity as simply the \emph{average information
  gain}.

\subsection{Two spin-1/2 systems} 

The simplest example of relative parameter estimation arises in the
context of a pair of spin-1/2 systems. Alice prepares the product
state $|\mathbf{n}_{1}\rangle \otimes |\mathbf{n}_{2}\rangle$, where
$\left| \mathbf{n}\right\rangle$ is the eigenstate of
$\mathbf{J}\cdot\mathbf{n}$ with positive eigenvalue (note that every
state of a spin-1/2 system is an SU(2) coherent state). Bob's task is
to estimate the relative angle $\alpha =
\cos^{-1}(\mathbf{n}_{1}\cdot\mathbf{n}_{2})$ given no knowledge of
the collective orientation of the state.  Because the joint Hilbert
space decomposes into a $J=0$ and a $J=1$ irrep, the optimal POVM has
the form $\{\Pi _{A},\Pi _{S}\}$, where $\Pi _{A}=|\Psi ^{-}\rangle
\langle \Psi ^{-}|$ is the projector onto the antisymmetric $(J=0)$
subspace and $\Pi _{S}=\mathbb{I}-\Pi _{A}$ is the projector onto the
symmetric $(J=1)$ subspace. The conditional probability of outcomes
$A$ and $S$ given $\alpha$ are simply 
\begin{align}
  p(A|\alpha ) &= \mathrm{Tr}(\Pi_{A} \rho _{\alpha})
  =\tfrac{1}{2}\sin ^{2}(\alpha /2) \,, \nonumber \\
  p(S|\alpha) &= 1-p(A|\alpha ) \,.
\end{align}
The average information gain and the optimal guess for the value of
$\alpha$ depend on Bob's prior over $\alpha$.  We consider two natural
choices of prior.

\subsubsection{Parallel versus anti-parallel spins}  

This situation corresponds to a prior
$p(\alpha{=}0)=p(\alpha{=}\pi)=1/2$, yielding $p(A)=1/4$, $p(S)=3/4$
and posteriors 
\begin{align}
  p(\alpha{=}0|A)&=0\,,\quad  &p(\alpha{=}\pi |A) &=1\,, \nonumber \\
  p(\alpha{=}0|S)&=2/3\,, &p(\alpha{=}\pi|S) &=1/3\,.
\end{align}
Upon obtaining the antisymmetric outcome, Bob knows that the spins
were anti-parallel, whereas upon obtaining the symmetric outcome, they
are deemed to be twice as likely to have been parallel than
anti-parallel.  We find 
\begin{equation}
  I_{A}=1\,, \qquad I_{S}=\frac{5}{3}-\log _{2}3\simeq .08\,170 \,, 
\end{equation}
i.e.\ 1 bit of information is gained upon obtaining the
antisymmetric outcome, and 0.08170 bits for the symmetric outcome. On
average, Bob gains
$I_{\mathrm{av}}=\frac{1}{4}I_{A}+\frac{3}{4}I_{S}\simeq 0.3113$ bits
of information.

\subsubsection{Uniform prior for each system's spin direction}  

In this case, the prior over $\alpha $ is $p(\alpha )= \frac{1}{2}\sin
\alpha$.  This implies posteriors 
\begin{align}
  p(\alpha |A) &= \sin ^{2}(\alpha
  /2)\sin \alpha\,, \nonumber \\
  p(\alpha |S) &= \frac{1}{3}(2-\sin ^{2}(\alpha
  /2))\sin \alpha \,, 
\end{align}
which are peaked at $2\pi /3$ and $0.4094\pi$ respectively. It follows
that these are the best guesses for the angle $\alpha$ given each
possible outcome. Using the posteriors, we find $I_{A}\simeq 0.2786$,
$I_{S}\simeq 0.02702$, which yields $I_{\mathrm{av} }\simeq 0.08993$.
Less information is acquired than in the parallel-antiparallel
estimation problem, because angles near $\pi/2$ are more difficult to
distinguish.

\begin{figure}
\includegraphics[width=3.5in]{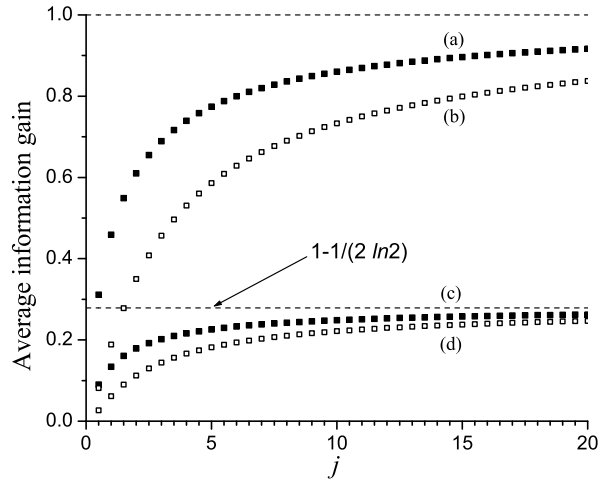}
\caption{Average information gain for measurements on a spin-1/2
  system and a spin-$j$ system.  The curves (a),(b) correspond to the
  optimal measurement and the optimal local measurement for the case
  when the spins are prepared parallel or antiparallel with equal
  probability.  The curves (c),(d) correspond to the optimal
  measurement and the optimal local measurements for the case when the
  initial direction of each spin is chosen uniformly from the sphere.}
\label{fig:Plot}
\end{figure}

\subsection{One spin-1/2, one spin-$j$ system} 

We now consider the estimation of the angle between a spin-1/2 system
and a spin-$j$ system for some arbitrary $j$, where the latter is in
an SU(2) coherent state $|j\mathbf{n}\rangle$ (the eigenstate of
$\mathbf{J}\cdot \mathbf{n}$ associated with the maximum
eigenvalue)~\cite{Arr72}.  Alice prepares $|\mathbf{n}_{1}\rangle
\otimes |j\mathbf{n}_{2}\rangle$ and Bob seeks to estimate $\alpha =
\cos^{-1}(\mathbf{n}_{1}\cdot\mathbf{n}_{2})$.  The joint Hilbert
space decomposes into a sum of a $J=j+1/2$ irrep and a $J=j-1/2$
irrep.  The optimal measurement is the two outcome POVM $\{ \Pi_+,
\Pi_- \}$, where $\Pi_{\pm}$ is the projector onto the $j\pm 1/2$
irrep~\footnote{This measurement is identical to the one described
  in~\cite{Fiu02} for optimal programmable measurements.}.  Using
Clebsch-Gordon coefficients, the probabilities for each of the
outcomes are found to be 
\begin{align}
  p(-|\alpha) &= {\rm Tr}(\Pi_- \rho_\alpha) =
  \tfrac{2j}{2j+1} \sin^2 (\alpha/2) \, , \nonumber \\
  p(+|\alpha) &= 1-p(-|\alpha) \,.
\end{align}
We again consider two possible priors over $\alpha$.

\subsubsection{Parallel versus anti-parallel spins}

A calculation similar to the one for two spin-1/2 systems yields the
posteriors 
\begin{align}
  p(\alpha{=}0|+)&=\frac{2j+1}{2j+2}\,, \quad  &p(\alpha{=}\pi
  |+) &= \frac{1}{2j+2} \,, \nonumber \\ 
  p(\alpha{=}0|-)&=0\,, &p(\alpha{=}\pi|-)&=1 \,.
\end{align}
Using these, we can calculate the average information gain as a
function of $j$; the result is curve~(a) of Fig.~\ref{fig:Plot}. The
$j=1/2$ value is the average information gain for two spin-1/2
systems, derived previously.  In the limit $j\to\infty$,
$p(\alpha{=}0|+)\to 1$ and $p(\alpha{=}\pi |+) \to 0$ so that the
outcome of the measurement leaves no uncertainty about whether the
spins were parallel or antiparallel, and the average information gain
goes to one bit.  Thus, in the limit that one of the spins becomes
large, the problem becomes equivalent to estimating whether the
spin-1/2 is up or down compared to some classical reference direction,
where one expects an average information gain of one bit.

\subsubsection{Uniform prior for each system's spin direction}

Following the same steps as before, the average information gain can
be derived as a function of $j$; the result is curve~(c) of
Fig.~\ref{fig:Plot}.  In the limit $j\to\infty$, we find $I_{\rm av}
= 1 - (2\ln 2)^{-1} \simeq 0.2787$ bits, which is precisely the
information gain for the optimal measurement of the angle of a
spin-1/2 system relative to a classical direction given a uniform
prior over spin directions~\cite{Per95}.

\subsection{Optimal local measurements} 

Consider again the simplest case of a pair of spin-1/2 systems.  The
optimal measurement in this case was found to be the POVM
$\{\Pi_A,\Pi_S\}$.  This measurement cannot be implemented by local
operations on the individual systems because $\Pi_A$ is a projector
onto an entangled state. We now determine the optimal local
measurement. We do so by first finding the optimal separable POVM (one
for which all the elements are separable operators), and then showing
that this can be achieved by LOCC.  (Not all separable POVMs can be
implemented using LOCC~\cite{Ben99}.)  The rotationally invariant
states for a pair of spin-1/2 systems, called Werner
states~\cite{Wer89}, have the form 
\begin{equation}
  \label{eq:WernerState}
  \rho_W=p\Pi_A+(1-p)\Pi_S/3 \,,
\end{equation}
and are known to only be separable for $p\leq 1/2$~\cite{Dur00}.  Thus,
the greatest relative weight of $\Pi_A$ to $\Pi_S$ that can occur in a
separable positive operator is $3$. The closest separable POVM to the
optimal POVM $\{\Pi_A,\Pi_S\}$ is therefore
$\{\Pi_A+\frac{1}{3}\Pi_S,\frac{2}{3}\Pi_S\}$.  We note that this POVM
is informationally equivalent to measuring the spin of each system
along the same (arbitrary) axis and registering whether the outcomes
are the same or not, which clearly only involves local operations (and
does not even require classical communication).  Because the POVM
$\{\Pi_A+\frac{1}{3}\Pi_S,\frac{2}{3}\Pi_S\}$ can be obtained by
random sampling of the outcome of $\{\Pi_A,\Pi_S\}$, the former is
strictly less informative than the latter. Indeed, the maximum average
information gain with the optimal local measurement is $0.0817$ bits
for case \emph{(1)} above, and $0.02702$ bits for case \emph{(2)},
both strictly less than those obtained for the optimal (joint)
measurement.

We extend this analysis to the spin-1/2, spin-$j$ case. Consider the
following LOCC measurement. The spin-$j$ system is measured along the
complete basis of SU(2) coherent states $\{|j\mathbf{n}_m\rangle \}_m$
where $m=0,\ldots,2j$ and $\mathbf{n}_m$ points at an angle $\theta_m
= \frac{2\pi m}{2j+1}$ in some fixed but arbitrary plane.  Then,
conditional on the outcome $m$ of this measurement, the spin-1/2
system is measured along the basis $\{|\mathbf{n}_m\rangle,
|{-}\mathbf{n}_m\rangle \}$.  The measurement outcome of the spin-$j$
system is then discarded, and all that is registered is whether the
outcome for the spin-1/2 system is $\pm \mathbf{n}_m$; i.e., whether
the two spins are aligned or anti-aligned.  The resulting 2-outcome
measurement is informationally equivalent to the rotationally
invariant POVM 
\begin{align}
  \Pi_1 &= \frac{2j+1}{2j+2} \Pi_+ \,, \nonumber \\
  \Pi_2 &= \Pi_-+\frac{1}{2j+2}\Pi_+ \,.
\end{align}
By numerically calculating the partial transpose of the operator
$\Pi_- + x\Pi_+$, the negativity of which is a necessary condition for
non-separability~\cite{Per96}, we find that $\{\Pi_1,\Pi_2\}$ is the
optimal separable POVM.  Thus, again, the optimal separable POVM can
be implemented by LOCC and gives less information than the optimal
(joint) measurement. The average information gain achieved by this
measurement, as a function of $j$, in cases \emph{(1)} and \emph{(2)}
are plotted as curves (b) and (d) of Fig.~\ref{fig:Plot}.  Note that
the optimum information gain overall \emph{can} be achieved by LOCC in
the limit $j \to \infty$.

\section{Discussion}

We now briefly discuss some other relative parameter estimation tasks
for which our result provides the solution.  The case we have yet to
address is the estimation of the angle between a spin-$j_1$ and a
spin-$j_2$ system, both in SU(2) coherent states, for arbitrary $j_1$,
$j_2$. Assuming $j_2\ge j_1$, the optimal measurement is the
$(2j_1+1)$-element projective measurement which projects onto the
subspaces of fixed total angular momentum $J$.  The posterior
distributions over $\alpha$ and the average information gain can be
calculated as before, although in this case they are much more
complicated.  However, in the limit $j_2 \to \infty$, the
Clebsch-Gordon coefficients simplify, and one can show that the
probability of a measurement outcome $J$ approaches the probabilities
obtained using the Born rule for a projective measurement along the
classical direction defined by the spin-$j_2$ system.  Thus,
the posterior distribution for any measurement result will agree with
what would be obtained classically, regardless of the prior over
$\alpha$.  If, in addition, we take $j_1\to\infty$, the information
gain for $\alpha$ becomes infinite (for any prior distribution) and
thus $\alpha$ can be inferred with certainty from the measurement
result, as expected for a measurement of the angle between two
classical directions.  Our results also indicate that, in the
classical limit, a measurement of the magnitude of total angular
momentum should be sufficient to estimate the relative angle, which is
indeed the case if the magnitude of each spin is known.

It should be noted that estimating the relative angle between a pair
of SU(2) coherent states is of particular importance because
estimating the eccentricity of an elliptic Rydberg state of a Hydrogen
atom is an instance of the same problem~\cite{Lin03}.  Rydberg states
are significant because they can be prepared experimentally.  Our
results imply that an optimal estimation of eccentricity is in fact
straightforward to achieve experimentally because it involves only a
measurement of the magnitude of the total angular momentum of the
atom.

Our results are also applicable to systems other than spin.  For
example, for \emph{any} realization of a pair of 2-level systems
(qubits), the degree of nonorthogonality between their states
(measured by, say, the overlap $|\langle\psi_1|\psi_2\rangle|$) is
invariant under collective transformations and is thus a relative
parameter.  Our measurement is thus optimal for estimating this
nonorthogonality.

In addition to solving various estimation problems, we have shown that
a macroscopic spin in the appropriate limit is equivalent to a
classical external RF as far as relative parameter estimation is
concerned.  This result suggests that it may be possible to express all
measurements (and possibly all operations) in a covariant, relative
framework that respects the underlying symmetries of the theory.  Such
a framework is necessary if one wishes to abide by the principle,
which has been so fruitful in the study of space and time but has yet
to be embraced in the quantum context, that all degrees of freedom
must be defined in terms of relations.

There remain many important questions for future investigation.  While
we have focussed on estimating relative parameters of product states,
one can also consider relative parameters of entangled states, and
here the landscape becomes much richer. For instance, for a pair of
spin-1/2 systems, while the set of product states supports a single
relative parameter, the set of all two-qubit states supports three:
the angle between the spins in a term of the Schmidt
decomposition~\cite{Per95}, the phase between the two terms of this
decomposition, and the degree of entanglement.  Our measurement scheme
is optimal for estimating these relative parameters as well.  Given
the significance of entanglement for quantum information theory, there
is likely much to be learned from investigating other sorts of
relative quantum information.

\begin{acknowledgments}
  S.D.B.\ is partially supported by the QUPRODIS project through the
  European Union and the Commonwealth Government of Australia.  T.R.\ 
  is supported by the NSA \& ARO under contract No.\ DAAG55-98-C-0040,
  and the U.K. EPSRC.  R.W.S.\ is supported by NSERC of Canada.  We
  acknowledge the help of Jens Eisert, who contributed significantly
  to the solution of the optimal estimation problem, along with
  helpful discussions with Ignacio Cirac and his group at MPQ, Netanel
  Lindner and Michael Nielsen.
\end{acknowledgments}

\end{document}